\documentclass[
]{ceurart}
\usepackage{listings}
\begin{document}

\copyrightyear{2026}
\copyrightclause{Copyright for this paper by its authors.
  Use permitted under Creative Commons License Attribution 4.0
  International (CC BY 4.0).}

\conference{Shaping Future Human Connection: Social Augmentation through XR Technologies. In Extended Abstracts of the 2026 CHI Conference on Human Factors in Computing Systems (CHI EA ’26), April 13--17, 2026, Barcelona, Spain}

\title{Making the Invisible Visible: Toward Micro-Expression Visualization for Empathy in Social Interaction}

\author[1]{Feiyang Yin}[%
orcid=0009-0008-3829-9200,
email=yin.feiyang.yj9@naist.ac.jp,
]
\address[1]{Nara Institute of Science and Technology, 8916-5, Takayama-cho, Ikoma, Nara, Japan}

\author[1]{Isidro {Butaslac}~III}[
orcid=0000-0003-1172-7611,
email=isidro.b@naist.ac.jp,
]
\cormark[1]

\author[2]{Patrick Gebhard}[
orcid=0000-0002-5566-4520,
email=patrick.gebhard@dfki.de,
]
\address[2]{German Research Center for Artificial Intelligence (DFKI), Stuhlsatzenhausweg 3, 66123 Saarbrücken, Germany}

\author[1]{Monica Perusquía-Hernández}[
orcid=0000-0002-0486-1743,
email=m.perusquia@is.naist.jp,
]

\author[3]{Zhaofeng Niu}[
orcid=0000-0001-9894-4584,
email=zhaofengniu@qfnu.edu.cn,
]
\address[3]{Qufu Normal University, 57 Jingxuan West Road, Qufu, Shandong 273165, China}

\author[1]{Taishi Sawabe}[
orcid=0000-0001-9244-479X,
email=t.sawabe@is.naist.jp,
]

\author[1]{Hirokazu Kato}[
orcid=0000-0003-3921-2871,
email=kato@is.naist.jp,
]

\cortext[1]{Corresponding author.}

\begin{abstract}
  Micro-expressions are brief and subtle facial movements that convey nuanced affective information but often remain imperceptible during natural social interaction. Although prior research has primarily focused on computational recognition and spotting of micro-expressions, their application in human-centered contexts remains limited. From the perspective of social augmentation, this work proposes a conceptual framework for micro-expression visualization that transforms otherwise imperceptible micro-expressions into perceptible affective cues, with the aim of exploring their potential influence on empathic experience. Furthermore, we outline a planned pilot study to preliminarily assess the feasibility of this framework under controlled conditions.
\end{abstract}


\maketitle

\section{Introduction}

Empathy is commonly described as the capacity to take another person’s perspective, enabling one to understand, feel, and respond to another’s experiences. As a fundamental component of social life, empathy plays a central role in everyday interpersonal interactions and emotional communication, shaping how individuals connect with others and navigate social relationships \cite{cuff_empathy_2016}. Beyond explicit reasoning or verbal exchange, empathy is also closely tied to how people perceive and make sense of others’ emotional experiences in social interaction. In this sense, the ability to perceive another person’s emotional states can be considered a crucial aspect of empathy \cite{decety_functional_2004}.

Facial expressions serve as a fundamental channel for affective perception during social interaction \cite{ekman_constants_1971, mehrabian_silent_1971, ekman_argument_1992}. As visible and shared signals, facial expressions provide immediate information about others’ emotional states and play an important role in supporting empathic understanding during face-to-face communication \cite{frith_role_2009}. Moreover, by eliciting unconscious facial mimicry, perceiving others’ facial expressions supports empathic responses in observers \cite{dimberg_facial_1982, hatfield_emotional_1993, dimberg_emotional_2011}. However, not all affective information expressed on the face is equally visible or easy to notice in natural social interaction.

Facial expressions can be divided into macro-expressions (MaEs, also known as regular or standard expressions) and micro-expressions (MEs) \cite{ekman_nonverbal_1969, yu_lgsnet_2024}. MaEs typically last between 0.5 and 4 seconds and are expressed with high intensity, making them consciously perceptible and highly salient in everyday social interaction \cite{ekman_nonverbal_1969, frank_behavioral_1993}. In contrast, MEs are brief and subtle facial movements that usually last less than 0.5 seconds and occur with low intensity, making them difficult to perceive during natural social interaction \cite{yan_how_2013}. Importantly, prior studies suggest that MEs may reflect underlying emotional states that are less subject to deliberate control, as they tend to be elicited automatically, even without conscious awareness of the emotional stimulus \cite{ekman_nonverbal_1969, gottschalk_micromomentary_1966}. Nevertheless, despite their potential to convey critical affective information, the fleeting and subtle nature of MEs renders them largely imperceptible in natural social interaction.

\begin{figure}
  \centering
  \includegraphics[width=0.5\linewidth]{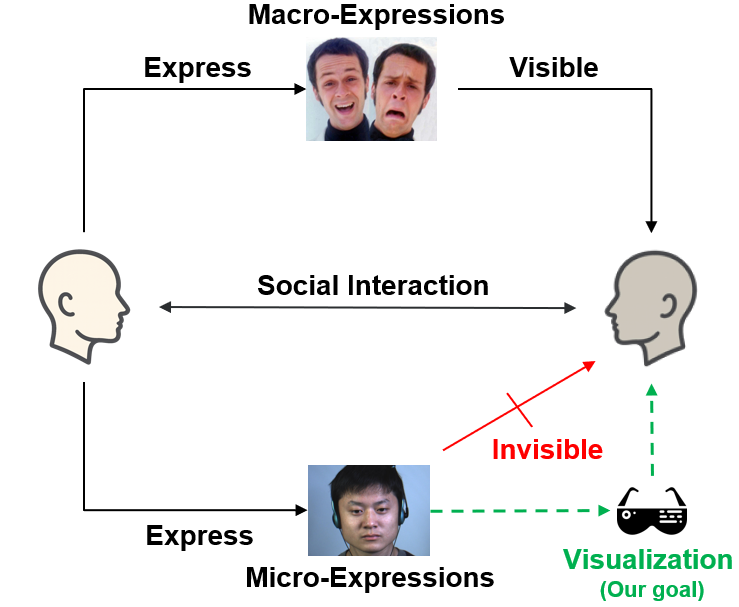}
  \caption{
    Overview of our Research Motivation.
    \newline
    This figure shows two people engaged in social interaction. The left person shows facial expressions. Macro-expressions are shown as visible and perceived by the observer, while micro-expressions are shown as invisible and unperceived. A visualization component illustrates the goal of making micro-expressions perceptible as affective cues to the observer.
    \newline (Top image source: simple.wikipedia.org/wiki/Facial\_expression, by Southerly Clubs, license: CC BY-SA 3.0)
    \newline (Bottom image source: reproduced from Yan et al. \cite{yan_casme_2014})
    }
  \label{fig:overview}
\end{figure}

As a consequence, observers may miss critical affective cues that are potentially important for empathic understanding. In this work, we aim to make MEs perceptible through visualization, as illustrated in Fig. \ref{fig:overview}. We also aim to investigate how transforming such affective cues from invisible (i.e., difficult to perceive with the unaided/untrained human eye) to visible may influence empathy in social interaction. Here, empathy is treated as an integration of cognitive empathy and affective empathy.

\section{Related Work}

In this section, we review prior studies on visible facial expressions and empathy, psychological research on MEs and empathy, advances in MEs spotting and recognition, and recent work on social augmentation for empathy support. Together, these lines of work situate our research vision and motivate our pilot study.

\subsection{Visible Facial Expressions and Empathy}

According to foundational work in nonverbal communication, when verbal content and nonverbal cues convey inconsistent information, observers tend to rely more heavily on nonverbal signals (particularly facial expressions) when interpreting a speaker’s emotional states \cite{mehrabian_silent_1971}. This reliance highlights the central role of visible facial expressions as a primary source of affective information in empathic understanding.

Early electromyography studies demonstrated that exposure to emotional facial expressions, such as happiness or anger, reliably induces congruent facial muscle activity in observers, even in the absence of conscious intention \cite{dimberg_facial_1982, dimberg_unconscious_2000}. These findings revealed that facial responses to others’ expressions can occur automatically and provided the empirical basis for the concept of facial mimicry, which describes the unconscious tendency to mirror others’ facial movements \cite{dimberg_facial_1982, dimberg_unconscious_2000, niedenthal_embodying_2007}. Further evidence from studies of neonates shows that even newborn infants exhibit spontaneous imitation of facial gestures, suggesting that facial mimicry emerges early in life and does not depend on explicit training or social learning \cite{meltzoff_imitation_1977}.

Through facial feedback processes, these automatically elicited muscular responses can evoke corresponding emotional states in the observer, a phenomenon commonly referred to as emotional contagion \cite{hatfield_emotional_1993, hatfield_new_2014}. Accordingly, effective perception of visible facial cues is essential for activating this innate internal simulation process, which supports empathic responses between individuals.

\subsection{Micro-Expressions and Empathy}

While MaEs are socially prominent, MEs offer a more nuanced window into an individual's emotional experience. Their emergence is widely regarded as a critical cue for understanding subtle emotional processes \cite{ekman_nonverbal_1969, ekman_darwin_2003}.

However, the fleeting and subtle nature of MEs creates a significant perceptual barrier that compromises the empathic process described in the previous section. Research suggests that MEs last less than 0.5 seconds, falling below the threshold for conscious detection by untrained observers \cite{ekman_nonverbal_1969, frank_behavioral_1993, matsumoto_evidence_2011}. 
Consequently, these critical affective cues often remain unperceived during social interaction, resulting in the loss of potentially valuable affective information \cite{perusquia-hernandez_invisible_2019}. The gap between the presence of affective information and the observer's limited ability to perceive it constrains empathic understanding during natural social interaction.

Despite these perceptual constraints, evidence from studies using the Micro Expression Training Tool demonstrates that enhancing sensitivity to MEs significantly improves empathic understanding, thereby fostering better rapport in clinical and interpersonal settings \cite{ekman_darwin_2003, matsumoto_evidence_2011, endres_micro-expression_2009}. These findings suggest that the affective information contained in MEs is vital for fostering empathy, but its impact remains limited by the constraints of human perception. This limitation provides a compelling motivation for technological interventions aimed at ME visualization, thereby improving the observer's empathy during social interaction.

\subsection{Micro-Expression Annotation and Analysis}

Micro-expression analysis is generally divided into two tasks: ME spotting and recognition. Research on ME analysis has traditionally relied on carefully annotated datasets, with annotation typically performed using the Facial Action Coding System (FACS). This process requires certified professional coders to identify subtle facial muscle movements, known as Action Units (AUs), through frame-by-frame analysis \cite{ekman_facial_1978, ekman_what_2005, zhao_facial_2023, dong_spontaneous_2022}. As a result, ME annotation is time-consuming, costly, and difficult to scale, leading to relatively small datasets collected primarily under controlled conditions \cite{yan_casme_2014, ekman_facial_1978}.

In the early stages of the field, ME recognition research focused predominantly on physiological signals, such as electroencephalography and electromyography, to capture internal affective changes that precede or accompany facial affective information \cite{tassinary_unobservable_1992, picard_affective_1997}. With the rapid development of deep learning, the field has increasingly shifted toward computer-vision-based approaches \cite{shu_review_2018, oh_survey_2018}. By leveraging advanced architectures, recent methods have achieved substantially improved recognition performance on benchmark datasets including CASME II, SAMM, and SMIC \cite{yan_casme_2014, davison_samm_2018, li_spontaneous_2013}. Building on these advances, multimodal recognition approaches that integrate facial video with physiological or thermal signals have further enhanced classification robustness \cite{saffaryazdi_using_2022}. In parallel, multimodal ME datasets have begun to emerge, providing new resources for ME analysis \cite{ma_mmme_2026}. Together, these developments indicate that once an ME is accurately segmented, its emotional category can be classified with relatively high reliability.

In contrast, ME spotting (i.e., identifying the temporal locations of MEs within long and unsegmented video sequences) remains considerably more challenging \cite{jain_micro-expressions_2024}. Spotting is inherently closer to real-world scenarios, where MEs occur sparsely amid neutral or unrelated facial movements. Although recent methods have reported progress, spotting accuracy in spontaneous and unconstrained settings remains limited, with persistent issues such as false positives and temporal imprecision \cite{yang_ofct_2025, guo_boosting_2025}. Consequently, while advances in ME recognition have made ME analysis increasingly feasible, reliable spotting continues to represent a critical bottleneck for practical deployment.

\subsection{Social Augmentation for Empathy}

Despite substantial progress in ME recognition and spotting, much of the existing work remains focused on algorithmic analysis and evaluation, and has yet to be applied in real-world settings. To transition these technical advancements into human-centered environments, a promising approach is to leverage these advancements for social augmentation. Social augmentation seeks to use computationally mediated elements to externalize or amplify latent affective signals, thereby supporting human social perception and interpersonal understanding \cite{damian_augmenting_2015, roth_technologies_2019, raisamo_human_2019}.

Prior studies in this space have explored various forms of social augmentation, such as providing users with additional affective or behavioral cues to facilitate the interpretation of others’ emotional states and social intentions \cite{damian_augmenting_2015, rojas_towards_2022}. Rather than replacing human judgment, these approaches aim to complement natural social perception by making otherwise implicit information more accessible.

The rapid development of eXtended Reality (XR) has recently provided a powerful vehicle for such interventions \cite{wang_raise_2025, lacle-melendez_virtual_2025}. However, existing research combining XR and MEs has largely been conducted in Virtual Reality (VR), where MEs are embedded in digital avatars to investigate their impact on users’ empathy in fully simulated environments \cite{tastemirova_microexpressions_2022, della_greca_user_2024, montanha_micro_2025}. While these virtual-world studies validate the potential of ME-based feedback, they often bypass the complexities of physical presence and spontaneous human encounters. In contrast, Augmented Reality (AR) is particularly well suited for social augmentation, as it blends virtual information with the physical environment \cite{valente_empathic_2022, yoneyama_augmented_2025}. This unique advantage makes AR an ideal platform for ME visualization, allowing subtle affective cues to be seamlessly overlaid onto the user’s real-world field of view during social interaction.

\section{Research Questions}

Despite existing research suggesting that MEs convey affective information relevant to empathy, their low intensity and short duration make them largely imperceptible during natural social interaction. 
As a result, it remains unclear whether making such latent affective cues perceptible through visualization can meaningfully influence empathic experience, and how different visualization strategies compare in their effectiveness. Guided by this gap, we propose the following research questions:

\textbf{RQ1:} \textit{Does visualizing micro-expressions influence observers' empathic experience during social interaction?}

\textbf{RQ2:} \textit{How do different micro-expression visualization strategies, with varying levels of abstraction and intervention, compare in their effectiveness for supporting empathic experience?}

\section{Conceptual Framework}

To address the challenges of ME perception, we propose a framework that converts such latent facial affective signals into visible cues, as depicted in Fig. \ref{fig:framework}. The framework is organized into three primary stages: facial video acquisition, ME analysis, and visualization of ME cues.

\begin{figure}[ht]
  \centering
  \includegraphics[width=0.85\linewidth]{framework_v3.png}
  \caption{
    Conceptual framework of the micro-expression (ME) visualization system.
    \newline
    During social interaction, a high-speed camera captures facial video streams of the observed person. The video stream is then transmitted to a server, where MEs are analyzed and ME cues are extracted, including inferred affective information and facial action units. These cues are subsequently sent to an AR device worn by the observer and visualized in real time to support perception during social interaction. 
    \newline
    The example on the right illustrates different forms of ME cues that may be visualized, including (from top to bottom): (a) text, (b) line-based motion cues, (c) micro-to-macro transformation.
    \newline (Image source: reproduced from Yan et al. \cite{yan_casme_2014})
    }
  \label{fig:framework}
\end{figure}

\subsection{Facial Video Acquisition}

In natural social interaction, MEs are embedded within continuous facial behavior and occur briefly and spontaneously \cite{ekman_nonverbal_1969, gottschalk_micromomentary_1966, feng_meldae_2025}. Accordingly, capturing MEs requires continuous facial video acquisition rather than isolated image sampling.

Beyond continuous capture, sufficient temporal resolution is also essential. Higher frame rates preserve fine-grained temporal information and enable the observation of rapid facial movements associated with MEs \cite{merghani_implication_2019}. Consequently, existing ME datasets and analysis pipelines are predominantly based on high-frame-rate recordings \cite{jain_micro-expressions_2024, merghani_implication_2019}. To remain consistent with these research practices and to support subsequent ME analysis, the proposed framework assumes high-frame-rate facial video acquisition.

\subsection{Micro-Expression Analysis}

Once facial video streams are received, the server performs ME analysis on the continuous input, which consists of two stages: ME spotting and ME recognition.

Micro-expression spotting aims to identify and extract candidate ME segments from the video stream. These extracted segments are then used as input for ME recognition, which analyzes each segment to infer ME cues, including facial action units and associated affective information. The resulting ME cues are subsequently transmitted to the AR device, where they are visualized to support perception during social interaction.

\subsection{Visualization of Micro-Expression Cues}

Given that ME visualization is intended to be used during ongoing face-to-face interaction, it is important that the visualization introduces minimal disruption to natural social dynamics. Accordingly, we plan to adopt an optical see-through AR device, which allows visual information to be overlaid onto the physical environment with relatively low intrusiveness, thereby preserving face-to-face interaction as much as possible \cite{bimber_modern_2005, nijholt_experiencing_2021}.

To investigate how different visualization strategies influence perception, it is essential that these approaches exhibit clear and meaningful contrasts. We therefore intend to explore three visualization strategies with deliberately different levels of abstraction and intervention. The first is a minimal textual visualization that explicitly presents the inferred emotional state. The second emphasizes ME cues at the physical level by enhancing facial muscle movements through line-based visual overlays derived from facial action units. The third operates at the perceptual level by transforming brief MEs into corresponding MaEs that reflect the same inferred emotional state.

\section{Pilot Study}

\begin{figure}[ht]
  \centering
  \includegraphics[width=0.85\linewidth]{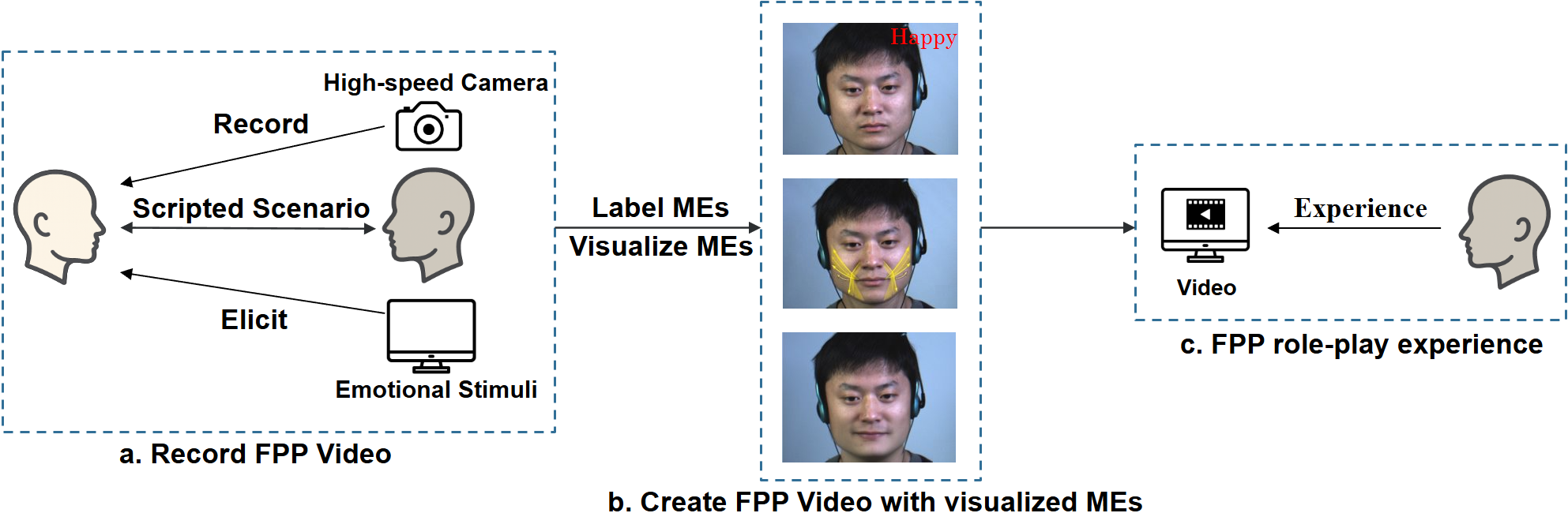}
  \caption{
    Workflow of the planned pilot study.
    \newline
    \emph{Fig. \ref{fig:pilot-workflow}a:} In a scripted scenario, participants are exposed to emotional stimuli while facial video is recorded using a high-speed camera from a first-person perspective (FPP). \emph{Fig. \ref{fig:pilot-workflow}b:} The recorded video is subsequently annotated to identify MEs and generate corresponding visualizations. Based on the labeled ME cues, augmented FPP videos are created using different visualization strategies. \emph{Fig. \ref{fig:pilot-workflow}c:} These videos are then presented to participants on a monitor to create an FPP role-play experience.
    \newline (Image source: reproduced from Yan et al. \cite{yan_casme_2014})
    }
  \label{fig:pilot-workflow}
\end{figure}

Although we have proposed a conceptual framework for the system, several sources of uncertainty remain when considering its deployment in natural conversational settings. First, MEs are inherently spontaneous and sparse, making them difficult to elicit and reproduce reliably in natural social interaction. In addition, despite ongoing progress in the ME analysis field, the accuracy of ME spotting in unconstrained scenarios remains limited, introducing uncertainty in reliable detection. Furthermore, existing ME recognition methods are primarily developed and evaluated using controlled datasets, and their performance in real-world environments may be affected by noise, occlusion, and interpersonal variability.

To address these uncertainties, Fig. \ref{fig:pilot-workflow} summarizes our pilot study design, which is conducted in a deliberately controlled setting to preliminarily explore how ME visualization may influence empathic experience and assess the plausibility of the proposed research vision.

\begin{figure}[ht]
  \centering
  \includegraphics[width=0.85\linewidth]{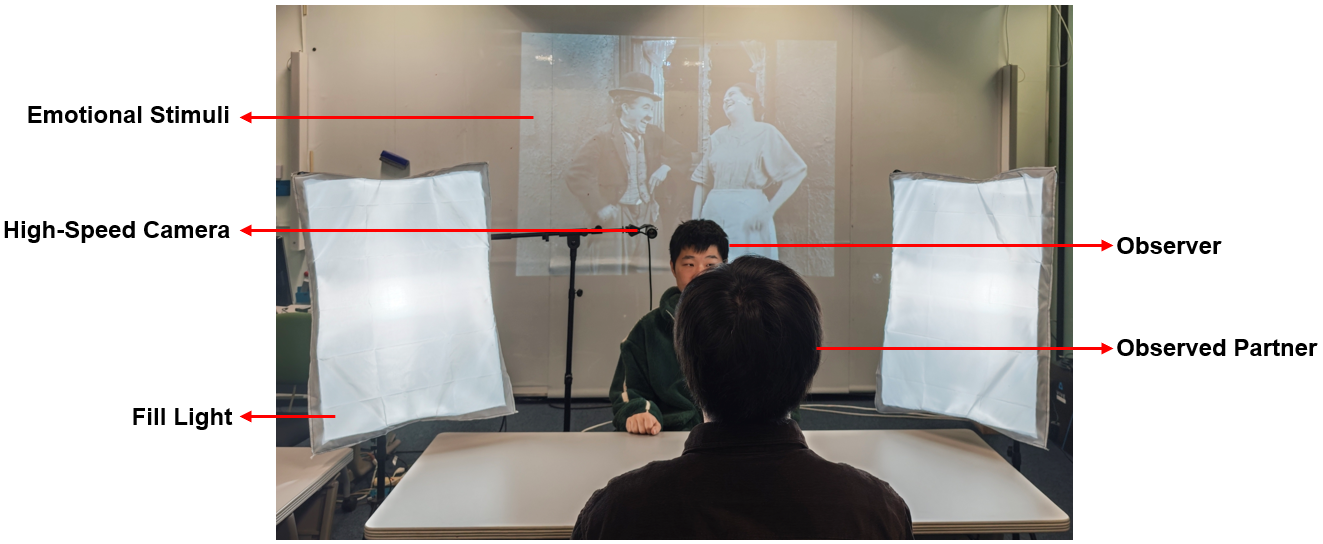}
  \caption{
    Recording environment for first-person perspective video.
    \newline
    During the recording stage, one participant serves as the observer and the other as the observed partner. As they engage in a scripted interaction, a high-speed camera will record from an observer's perspective. When the interaction reaches predefined moments in the script, emotional stimuli are introduced to elicit MEs from the observed partner.
    \newline (Background image source: screenshot from *The Kid* (1921), directed by Charlie Chaplin. Public domain.)
    }
  \label{fig:recording}
\end{figure}

First (Fig. \ref{fig:pilot-workflow}a), to address the difficulty of reliably eliciting MEs in natural social interaction, we designed scripted interaction scenarios combined with emotion-eliciting stimuli to induce MEs in a more controlled manner. These scenarios are intended to allow MEs to emerge spontaneously while improving their reproducibility across recordings. Our experimental setup for this recording environment is presented in Fig. \ref{fig:recording}. We also plan to collaborate with psychology experts to ensure the validity and effectiveness of both the interaction scripts and the emotion-eliciting stimuli.

Second (Fig. \ref{fig:pilot-workflow}b), to ensure the accuracy of ME cues and avoid uncertainties introduced by automatic analysis in unconstrained settings, recorded videos will be manually screened and annotated by two certified FACS coders. The coders identify ME segments and label associated facial action units through frame-by-frame analysis. Based on these verified annotations, multiple versions of the same videos are generated with different ME visualization strategies applied.

Finally (Fig. \ref{fig:pilot-workflow}c), participants experience these videos from a first-person perspective and compare different visualization conditions. Because the visualized cues are derived from verified manual annotations rather than real-time inference, the pilot study is able to isolate the perceptual and experiential effects of ME visualization itself, without being confounded by detection or recognition errors.

Through this pipeline, the pilot study serves as an initial feasibility exploration of the proposed research vision. By establishing a controlled foundation, it allows us to examine whether transforming previously imperceptible MEs into visible cues has the potential to influence empathic experience, while informing future work toward more ecologically valid and interactive settings. 

This study design was approved by the Ethics Committee of the Nara Institute of Science and Technology (2025-I-42).

\section{Discussion}

In this work, we presented a research vision and a conceptual framework for ME visualization as a form of social augmentation, and outlined the design of a pilot study to explore its feasibility. Rather than focusing on a specific application, our discussion centers on a generalized social interaction setting in which subtle affective cues are difficult to perceive yet potentially meaningful for empathic understanding.

This generalized framing leaves room for future exploration of more concrete application contexts. For example, ME visualization may support “reading the room” (i.e., sensing subtle social and emotional cues within a group) in everyday social interaction, assist clinicians in better understanding patients' emotional states, or provide complementary cues in investigative settings. Exploring such domain-specific scenarios will be an important direction for future work.

\section*{Declaration on Generative AI}
During the preparation of this work, the authors used ChatGPT in order to: generate images; translate text; paraphrase and reword sentences; improve writing style; and check grammar and spelling errors. After using this tool, the authors reviewed and edited the content as needed and take full responsibility for the publication's content.

\bibliography{my_library}

@article{frank_behavioral_1993,
	title = {Behavioral markers and recognizability of the smile of enjoyment},
	volume = {64},
	doi = {10.1037//0022-3514.64.1.83},
	number = {1},
	journal = {Journal of Personality and Social Psychology},
	author = {Frank, Mark G. and Ekman, Paul and Friesen, Wallace V.},
	month = jan,
	year = {1993},
	pages = {83--93},
}

@book{picard_affective_1997,
	address = {Cambridge, MA, US},
	title = {Affective computing},
	publisher = {The MIT Press},
	author = {Picard, Rosalind W.},
	year = {1997},
}

@misc{mehrabian_silent_1971,
	title = {Silent messages.},
	journal = {APA PsycNET},
	author = {Mehrabian, Albert},
	year = {1971},
}

@article{ma_mmme_2026,
	title = {{MMME}: {A} spontaneous multi-modal micro-expression dataset enabling visual-physiological fusion},
	volume = {129},
	shorttitle = {{MMME}},
	doi = {10.1016/j.inffus.2025.104038},
	journal = {Information Fusion},
	author = {Ma, Chuang and Pei, Yu and Zhang, Jianhang and Zhao, Shaokai and Ji, Bowen and Xie, Liang and Yan, Ye and Yin, Erwei},
	month = may,
	year = {2026},
	pages = {104038},
}

@inproceedings{bimber_modern_2005,
	address = {New York, NY, USA},
	series = {{SIGGRAPH} '05},
	title = {Modern approaches to augmented reality},
	doi = {10.1145/1198555.1198711},
	booktitle = {{ACM} {SIGGRAPH} 2005 {Courses}},
	publisher = {Association for Computing Machinery},
	author = {Bimber, Oliver and Raskar, Ramesh},
	year = {2005},
	pages = {1--es},
}

@inproceedings{perusquia-hernandez_invisible_2019,
	address = {New York, NY, USA},
	series = {{CHI} '19},
	title = {The {Invisible} {Potential} of {Facial} {Electromyography}: {A} {Comparison} of {EMG} and {Computer} {Vision} when {Distinguishing} {Posed} from {Spontaneous} {Smiles}},
	shorttitle = {The {Invisible} {Potential} of {Facial} {Electromyography}},
	doi = {10.1145/3290605.3300379},
	booktitle = {Proceedings of the 2019 {CHI} {Conference} on {Human} {Factors} in {Computing} {Systems}},
	publisher = {Association for Computing Machinery},
	author = {Perusquía-Hernández, Monica and Ayabe-Kanamura, Saho and Suzuki, Kenji and Kumano, Shiro},
	year = {2019},
	pages = {1--9},
}

@article{hatfield_emotional_1993,
	title = {Emotional {Contagion}},
	volume = {2},
	doi = {10.1111/1467-8721.ep10770953},
	number = {3},
	journal = {Current Directions in Psychological Science},
	publisher = {SAGE Publications Inc},
	author = {Hatfield, Elaine and Cacioppo, John T. and Rapson, Richard L.},
	month = jun,
	year = {1993},
	pages = {96--100},
}

@article{ekman_nonverbal_1969,
	title = {Nonverbal leakage and clues to deception},
	volume = {32},
	doi = {10.1080/00332747.1969.11023575},
	number = {1},
	journal = {Psychiatry},
	author = {Ekman, Paul and Friesen, Wallace. V.},
	month = feb,
	year = {1969},
	pages = {88--106},
}

@incollection{gottschalk_micromomentary_1966,
	address = {Boston, MA},
	title = {Micromomentary facial expressions as indicators of ego mechanisms in psychotherapy},
	doi = {10.1007/978-1-4684-6045-2_14},
	booktitle = {Methods of {Research} in {Psychotherapy}},
	publisher = {Springer US},
	author = {Gottschalk, Louis A. and Auerbach, Arthur H. and Haggard, Ernest A. and Isaacs, Kenneth S.},
	year = {1966},
	pages = {154--165},
}

@article{yang_ofct_2025,
	title = {Ofct: a micro-expression spotting method fusing optical flow features and category text information},
	volume = {11},
	shorttitle = {Ofct},
	doi = {10.1007/s40747-025-01997-4},
	number = {8},
	journal = {Complex \& Intelligent Systems},
	author = {Yang, Henian and Huang, Shucheng and Li, Mingxing},
	month = jul,
	year = {2025},
	pages = {373},
}

@article{montanha_micro_2025,
	title = {Micro and macro facial expressions by driven animations in realistic {Virtual} {Humans}},
	volume = {52},
	doi = {10.1016/j.entcom.2024.100853},
	journal = {Entertainment Computing},
	author = {Montanha, Rubens Halbig and Raupp, Giovana Nascimento and Schmitt, Ana Carolina Policarpo and Araujo, Victor Flávio de Andrade and Musse, Soraia Raupp},
	month = jan,
	year = {2025},
	pages = {100853},
}

@article{yu_lgsnet_2024,
	title = {{LGSNet}: {A} {Two}-{Stream} {Network} for {Micro}- and {Macro}-{Expression} {Spotting} {With} {Background} {Modeling}},
	volume = {15},
	shorttitle = {{LGSNet}},
	doi = {10.1109/TAFFC.2023.3266808},
	number = {1},
	journal = {IEEE Transactions on Affective Computing},
	author = {Yu, Wang-Wang and Jiang, Jingwen and Yang, Kai-Fu and Yan, Hong-Mei and Li, Yong-Jie},
	month = jan,
	year = {2024},
	pages = {223--240},
}

@article{yan_how_2013,
	title = {How {Fast} are the {Leaked} {Facial} {Expressions}: {The} {Duration} of {Micro}-{Expressions}},
	volume = {37},
	shorttitle = {How {Fast} are the {Leaked} {Facial} {Expressions}},
	doi = {10.1007/s10919-013-0159-8},
	number = {4},
	journal = {Journal of Nonverbal Behavior},
	author = {Yan, Wen-Jing and Wu, Qi and Liang, Jing and Chen, Yu-Hsin and Fu, Xiaolan},
	month = dec,
	year = {2013},
	pages = {217--230},
}

@article{ekman_argument_1992,
	title = {An argument for basic emotions},
	volume = {6},
	doi = {10.1080/02699939208411068},
	number = {3-4},
	journal = {Cognition and Emotion},
	publisher = {Routledge},
	author = {Ekman, Paul},
	month = may,
	year = {1992},
	pages = {169--200},
}

@article{decety_functional_2004,
	title = {The {Functional} {Architecture} of {Human} {Empathy}},
	volume = {3},
	doi = {10.1177/1534582304267187},
	number = {2},
	journal = {Behavioral and Cognitive Neuroscience Reviews},
	publisher = {SAGE Publications},
	author = {Decety, Jean and Jackson, Philip L.},
	month = jun,
	year = {2004},
	pages = {71--100},
}

@article{cuff_empathy_2016,
	title = {Empathy: {A} {Review} of the {Concept}},
	volume = {8},
	shorttitle = {Empathy},
	doi = {10.1177/1754073914558466},
	number = {2},
	journal = {Emotion Review},
	publisher = {SAGE Publications},
	author = {Cuff, Benjamin M.P. and Brown, Sarah J. and Taylor, Laura and Howat, Douglas J.},
	month = apr,
	year = {2016},
	pages = {144--153},
}

@misc{ekman_facial_1978,
	title = {Facial {Action} {Coding} {System}},
	doi = {10.1037/t27734-000},
	author = {Ekman, Paul and Friesen, Wallace V.},
	year = {1978},
}

@article{niedenthal_embodying_2007,
	title = {Embodying {Emotion}},
	volume = {316},
	doi = {10.1126/science.1136930},
	number = {5827},
	journal = {Science},
	publisher = {American Association for the Advancement of Science},
	author = {Niedenthal, Paula M.},
	month = may,
	year = {2007},
	pages = {1002--1005},
}

@inproceedings{damian_augmenting_2015,
	address = {New York, NY, USA},
	series = {{CHI} '15},
	title = {Augmenting {Social} {Interactions}: {Realtime} {Behavioural} {Feedback} using {Social} {Signal} {Processing} {Techniques}},
	shorttitle = {Augmenting {Social} {Interactions}},
	doi = {10.1145/2702123.2702314},
	booktitle = {Proceedings of the 33rd {Annual} {ACM} {Conference} on {Human} {Factors} in {Computing} {Systems}},
	publisher = {Association for Computing Machinery},
	author = {Damian, Ionut and Tan, Chiew Seng (Sean) and Baur, Tobias and Schöning, Johannes and Luyten, Kris and André, Elisabeth},
	year = {2015},
	pages = {565--574},
}

@inproceedings{roth_technologies_2019,
	address = {New York, NY, USA},
	series = {{VRST} '19},
	title = {Technologies for {Social} {Augmentations} in {User}-{Embodied} {Virtual} {Reality}},
	doi = {10.1145/3359996.3364269},
	booktitle = {Proceedings of the 25th {ACM} {Symposium} on {Virtual} {Reality} {Software} and {Technology}},
	publisher = {Association for Computing Machinery},
	author = {Roth, Daniel and Bente, Gary and Kullmann, Peter and Mal, David and Purps, Chris Felix and Vogeley, Kai and Latoschik, Marc Erich},
	year = {2019},
	pages = {1--12},
}

@misc{feng_meldae_2025,
	title = {{MELDAE}: {A} {Framework} for {Micro}-{Expression} {Spotting}, {Detection}, and {Automatic} {Evaluation} in {In}-the-{Wild} {Conversational} {Scenes}},
	shorttitle = {{MELDAE}},
	doi = {10.48550/arXiv.2510.22575},
	publisher = {arXiv},
	author = {Feng, Yigui and Wang, Qinglin and Liu, Yang and Liu, Ke and Mo, Haotian and Huang, Enhao and Liu, Gencheng and Liu, Mingzhe and Liu, Jie},
	month = oct,
	year = {2025},
}

@article{raisamo_human_2019,
	series = {50 years of the {International} {Journal} of {Human}-{Computer} {Studies}. {Reflections} on the past, present and future of human-centred technologies},
	title = {Human augmentation: {Past}, present and future},
	volume = {131},
	shorttitle = {Human augmentation},
	doi = {10.1016/j.ijhcs.2019.05.008},
	journal = {International Journal of Human-Computer Studies},
	author = {Raisamo, Roope and Rakkolainen, Ismo and Majaranta, Päivi and Salminen, Katri and Rantala, Jussi and Farooq, Ahmed},
	month = nov,
	year = {2019},
	pages = {131--143},
}

@inproceedings{nijholt_experiencing_2021,
	address = {New York, NY, USA},
	series = {{UbiComp}/{ISWC} '21 {Adjunct}},
	title = {Experiencing {Social} {Augmented} {Reality} in {Public} {Spaces}},
	doi = {10.1145/3460418.3480157},
	booktitle = {Adjunct {Proceedings} of the 2021 {ACM} {International} {Joint} {Conference} on {Pervasive} and {Ubiquitous} {Computing} and {Proceedings} of the 2021 {ACM} {International} {Symposium} on {Wearable} {Computers}},
	publisher = {Association for Computing Machinery},
	author = {Nijholt, Anton},
	year = {2021},
	pages = {570--574},
}

@article{oh_survey_2018,
	title = {A {Survey} of {Automatic} {Facial} {Micro}-{Expression} {Analysis}: {Databases}, {Methods}, and {Challenges}},
	volume = {9},
	shorttitle = {A {Survey} of {Automatic} {Facial} {Micro}-{Expression} {Analysis}},
	doi = {10.3389/fpsyg.2018.01128},
	journal = {Frontiers in Psychology},
	publisher = {Frontiers},
	author = {Oh, Yee-Hui and See, John and Le Ngo, Anh Cat and Phan, Raphael C.-W. and Baskaran, Vishnu M.},
	month = jul,
	year = {2018},
}

@article{shu_review_2018,
	title = {A {Review} of {Emotion} {Recognition} {Using} {Physiological} {Signals}},
	volume = {18},
	doi = {10.3390/s18072074},
	number = {7},
	journal = {Sensors (Basel, Switzerland)},
	author = {Shu, Lin and Xie, Jinyan and Yang, Mingyue and Li, Ziyi and Li, Zhenqi and Liao, Dan and Xu, Xiangmin and Yang, Xinyi},
	month = jun,
	year = {2018},
	pages = {2074},
}

@article{tassinary_unobservable_1992,
	title = {Unobservable {Facial} {Actions} and {Emotion}},
	volume = {3},
	doi = {10.1111/j.1467-9280.1992.tb00252.x},
	number = {1},
	journal = {Psychological Science},
	publisher = {SAGE Publications Inc},
	author = {Tassinary, Louis G. and Cacioppo, John T.},
	month = jan,
	year = {1992},
	pages = {28--33},
}

@article{dong_spontaneous_2022,
	title = {Spontaneous {Facial} {Expressions} and {Micro}-expressions {Coding}: {From} {Brain} to {Face}},
	volume = {12},
	shorttitle = {Spontaneous {Facial} {Expressions} and {Micro}-expressions {Coding}},
	doi = {10.3389/fpsyg.2021.784834},
	journal = {Frontiers in Psychology},
	publisher = {Frontiers},
	author = {Dong, Zizhao and Wang, Gang and Lu, Shaoyuan and Li, Jingting and Yan, Wenjing and Wang, Su-Jing},
	month = jan,
	year = {2022},
}

@book{ekman_what_2005,
	address = {New York, NY, US},
	series = {What the face reveals: {Basic} and applied studies of spontaneous expression using the facial action coding system ({FACS}), 2nd ed},
	title = {What the face reveals: {Basic} and applied studies of spontaneous expression using the facial action coding system ({FACS}), 2nd ed},
	shorttitle = {What the face reveals},
	doi = {10.1093/acprof:oso/9780195179644.001.0001},
	publisher = {Oxford University Press},
	editor = {Ekman, Paul and Rosenberg, Erika L.},
	year = {2005},
}

@article{endres_micro-expression_2009,
	title = {Micro-expression recognition training in medical students: a pilot study},
	volume = {9},
	shorttitle = {Micro-expression recognition training in medical students},
	doi = {10.1186/1472-6920-9-47},
	journal = {BMC medical education},
	author = {Endres, Jennifer and Laidlaw, Anita},
	month = jul,
	year = {2009},
	pages = {47},
}

@article{hatfield_new_2014,
	title = {New {Perspectives} on {Emotional} {Contagion}: {A} {Review} of {Classic} and {Recent} {Research} on {Facial} {Mimicry} and {Contagion}},
	volume = {8},
	copyright = {Copyright (c)},
	shorttitle = {New {Perspectives} on {Emotional} {Contagion}},
	doi = {10.5964/ijpr.v8i2.162},
	number = {2},
	journal = {Interpersona: An International Journal on Personal Relationships},
	author = {Hatfield, Elaine and Bensman, Lisamarie and Thornton, Paul D. and Rapson, Richard L.},
	month = dec,
	year = {2014},
	pages = {159--179},
}

@article{ekman_darwin_2003,
	title = {Darwin, {Deception}, and {Facial} {Expression}},
	volume = {1000},
	doi = {10.1196/annals.1280.010},
	number = {1},
	journal = {Annals of the New York Academy of Sciences},
	author = {Ekman, Paul},
	year = {2003},
	pages = {205--221},
}

@article{matsumoto_evidence_2011,
	title = {Evidence for training the ability to read microexpressions of emotion},
	volume = {35},
	doi = {10.1007/s11031-011-9212-2},
	number = {2},
	journal = {Motivation and Emotion},
	author = {Matsumoto, David and Hwang, Hyi Sung},
	month = jun,
	year = {2011},
	pages = {181--191},
}

@article{ekman_constants_1971,
	address = {US},
	title = {Constants across cultures in the face and emotion},
	volume = {17},
	doi = {10.1037/h0030377},
	number = {2},
	journal = {Journal of Personality and Social Psychology},
	publisher = {American Psychological Association},
	author = {Ekman, Paul and Friesen, Wallace V.},
	year = {1971},
	pages = {124--129},
}

@article{meltzoff_imitation_1977,
	title = {Imitation of {Facial} and {Manual} {Gestures} by {Human} {Neonates}},
	volume = {198},
	doi = {10.1126/science.198.4312.75},
	number = {4312},
	journal = {Science},
	publisher = {American Association for the Advancement of Science},
	author = {Meltzoff, Andrew N. and Moore, M. Keith},
	month = oct,
	year = {1977},
	pages = {75--78},
}

@article{dimberg_unconscious_2000,
	title = {Unconscious {Facial} {Reactions} to {Emotional} {Facial} {Expressions}},
	volume = {11},
	doi = {10.1111/1467-9280.00221},
	number = {1},
	journal = {Psychological Science},
	publisher = {SAGE Publications Inc},
	author = {Dimberg, Ulf and Thunberg, Monika and Elmehed, Kurt},
	month = jan,
	year = {2000},
	pages = {86--89},
}

@article{yoneyama_augmented_2025,
	title = {Augmented conversations: {AR} face filters for facilitating comfortable in-person interactions},
	volume = {19},
	shorttitle = {Augmented conversations},
	doi = {10.1007/s12193-024-00446-9},
	number = {1},
	journal = {Journal on Multimodal User Interfaces},
	author = {Yoneyama, Juri and Fujimoto, Yuichiro and Okazaki, Kosuke and Sawabe, Taishi and Kanbara, Masayuki and Kato, Hirokazu},
	month = mar,
	year = {2025},
	pages = {57--74},
}

@inproceedings{wang_raise_2025,
	address = {New York, NY, USA},
	series = {{CHI} '25},
	title = {Raise {Your} {Eyebrows} {Higher}: {Facilitating} {Emotional} {Communication} in {Social} {Virtual} {Reality} {Through} {Region}-{Specific} {Facial} {Expression} {Exaggeration}},
	shorttitle = {Raise {Your} {Eyebrows} {Higher}},
	doi = {10.1145/3706598.3713688},
	booktitle = {Proceedings of the 2025 {CHI} {Conference} on {Human} {Factors} in {Computing} {Systems}},
	publisher = {Association for Computing Machinery},
	author = {Wang, Xueyang and Zhao, Sheng and Wang, Yihe and Han, Howard Ziyu and Liu, Xinge and Yi, Xin and Tong, Xin and Li, Hewu},
	year = {2025},
	pages = {1--22},
}

@inproceedings{rojas_towards_2022,
	address = {New York, NY, USA},
	series = {{CHI} {EA} '22},
	title = {Towards {Enhancing} {Empathy} {Through} {Emotion} {Augmented} {Remote} {Communication}},
	doi = {10.1145/3491101.3519797},
	publisher = {Association for Computing Machinery},
	author = {Rojas, Camilo and Zuccarelli, Eugenio and Chin, Alexandra and Patekar, Gaurav and Esquivel, David and Maes, Pattie},
	year = {2022},
	pages = {1--9},
}

@article{yan_casme_2014,
	title = {{CASME} {II}: {An} {Improved} {Spontaneous} {Micro}-{Expression} {Database} and the {Baseline} {Evaluation}},
	volume = {9},
	shorttitle = {{CASME} {II}},
	doi = {10.1371/journal.pone.0086041},
	number = {1},
	journal = {PLOS ONE},
	publisher = {Public Library of Science},
	author = {Yan, Wen-Jing and Li, Xiaobai and Wang, Su-Jing and Zhao, Guoying and Liu, Yong-Jin and Chen, Yu-Hsin and Fu, Xiaolan},
	month = jan,
	year = {2014},
	pages = {e86041},
}

@inproceedings{li_spontaneous_2013,
	title = {A {Spontaneous} {Micro}-expression {Database}: {Inducement}, collection and baseline},
	shorttitle = {A {Spontaneous} {Micro}-expression {Database}},
	doi = {10.1109/FG.2013.6553717},
	booktitle = {2013 10th {IEEE} {International} {Conference} and {Workshops} on {Automatic} {Face} and {Gesture} {Recognition} ({FG})},
	author = {Li, Xiaobai and Pfister, Tomas and Huang, Xiaohua and Zhao, Guoying and Pietikäinen, Matti},
	month = apr,
	year = {2013},
	pages = {1--6},
}

@article{davison_samm_2018,
	title = {{SAMM}: {A} {Spontaneous} {Micro}-{Facial} {Movement} {Dataset}},
	volume = {9},
	shorttitle = {{SAMM}},
	doi = {10.1109/TAFFC.2016.2573832},
	number = {1},
	journal = {IEEE Transactions on Affective Computing},
	author = {Davison, Adrian K. and Lansley, Cliff and Costen, Nicholas and Tan, Kevin and Yap, Moi Hoon},
	month = jan,
	year = {2018},
	pages = {116--129},
}

@article{jain_micro-expressions_2024,
	title = {Micro-expressions: a survey},
	volume = {83},
	shorttitle = {Micro-expressions},
	doi = {10.1007/s11042-023-17313-6},
	number = {18},
	journal = {Multimedia Tools and Applications},
	author = {Jain, Ankita and Bhakta, Dhananjoy},
	month = may,
	year = {2024},
	pages = {53165--53200},
}

@article{zhao_facial_2023,
	title = {Facial {Micro}-{Expressions}: {An} {Overview}},
	volume = {111},
	shorttitle = {Facial {Micro}-{Expressions}},
	doi = {10.1109/JPROC.2023.3275192},
	number = {10},
	journal = {Proceedings of the IEEE},
	author = {Zhao, Guoying and Li, Xiaobai and Li, Yante and Pietikäinen, Matti},
	month = oct,
	year = {2023},
	pages = {1215--1235},
}

@article{merghani_implication_2019,
	title = {The implication of spatial temporal changes on facial micro-expression analysis},
	volume = {78},
	doi = {10.1007/s11042-019-7434-6},
	number = {15},
	journal = {Multimedia Tools and Applications},
	author = {Merghani, Walied and Davison, Adrian K. and Yap, Moi Hoon},
	month = aug,
	year = {2019},
	pages = {21613--21628},
}

@article{frith_role_2009,
	title = {Role of facial expressions in social interactions},
	volume = {364},
	doi = {10.1098/rstb.2009.0142},
	number = {1535},
	journal = {Philosophical Transactions of the Royal Society B: Biological Sciences},
	publisher = {Royal Society},
	author = {Frith, Chris},
	month = dec,
	year = {2009},
	pages = {3453--3458},
}

@article{dimberg_emotional_2011,
	title = {Emotional {Empathy} and {Facial} {Reactions} to {Facial} {Expressions}},
	volume = {25},
	doi = {10.1027/0269-8803/a000029},
	number = {1},
	journal = {Journal of Psychophysiology},
	publisher = {Hogrefe Publishing},
	author = {Dimberg, Ulf and Andréasson, Per and Thunberg, Monika},
	month = jan,
	year = {2011},
	pages = {26--31},
}

@article{tastemirova_microexpressions_2022,
	title = {Microexpressions in digital humans: perceived affect, sincerity, and trustworthiness},
	volume = {32},
	shorttitle = {Microexpressions in digital humans},
	doi = {10.1007/s12525-022-00563-x},
	number = {3},
	journal = {Electronic Markets},
	author = {Tastemirova, Aliya and Schneider, Johannes and Kruse, Leona Chandra and Heinzle, Simon and Brocke, Jan vom},
	month = sep,
	year = {2022},
	pages = {1603--1620},
}

@inproceedings{della_greca_user_2024,
	address = {New York, NY, USA},
	series = {{AVI} '24},
	title = {A user study on the relationship between empathy and facial-based emotion simulation in {Virtual} {Reality}},
	doi = {10.1145/3656650.3656691},
	booktitle = {Proceedings of the 2024 {International} {Conference} on {Advanced} {Visual} {Interfaces}},
	publisher = {Association for Computing Machinery},
	author = {Della Greca, Attilio and Ilaria, Amaro and Tucci, Cesare and Frugieri, Nicola and Tortora, Genoveffa},
	year = {2024},
	pages = {1--9},
}

@inproceedings{guo_boosting_2025,
	address = {New York, NY, USA},
	series = {{MM} '25},
	title = {Boosting {Micro}-{Expression} {Analysis} via {Prior}-{Guided} {Video}-{Level} {Regression}},
	doi = {10.1145/3746027.3762026},
	booktitle = {Proceedings of the 33rd {ACM} {International} {Conference} on {Multimedia}},
	publisher = {Association for Computing Machinery},
	author = {Guo, Zizheng and Zou, Bochao and Jia, Yinuo and Li, Xiangyu and Ma, Huimin},
	year = {2025},
	pages = {13964--13971},
}

@article{saffaryazdi_using_2022,
	title = {Using {Facial} {Micro}-{Expressions} in {Combination} {With} {EEG} and {Physiological} {Signals} for {Emotion} {Recognition}},
	volume = {13},
	doi = {10.3389/fpsyg.2022.864047},
	journal = {Frontiers in Psychology},
	publisher = {Frontiers},
	author = {Saffaryazdi, Nastaran and Wasim, Syed Talal and Dileep, Kuldeep and Nia, Alireza Farrokhi and Nanayakkara, Suranga and Broadbent, Elizabeth and Billinghurst, Mark},
	month = jun,
	year = {2022},
}

@article{lacle-melendez_virtual_2025,
	title = {Virtual and augmented reality to develop empathy: a systematic literature review},
	volume = {84},
	shorttitle = {Virtual and augmented reality to develop empathy},
	doi = {10.1007/s11042-024-19191-y},
	number = {11},
	journal = {Multimedia Tools and Applications},
	author = {Lacle-Melendez, Jose and Silva-Medina, Sofia and Bacca-Acosta, Jorge},
	month = mar,
	year = {2025},
	pages = {8893--8927},
}

@inproceedings{valente_empathic_2022,
	title = {Empathic {AuRea}: {Exploring} the {Effects} of an {Augmented} {Reality} {Cue} for {Emotional} {Sharing} {Across} {Three} {Face}-to-{Face} {Tasks}},
	shorttitle = {Empathic {AuRea}},
	doi = {10.1109/VR51125.2022.00034},
	booktitle = {2022 {IEEE} {Conference} on {Virtual} {Reality} and {3D} {User} {Interfaces} ({VR})},
	author = {Valente, Andreia and Lopes, Daniel Simões and Nunes, Nuno and Esteves, Augusto},
	month = mar,
	year = {2022},
	pages = {158--166},
}

@article{dimberg_facial_1982,
	title = {Facial {Reactions} to {Facial} {Expressions}},
	volume = {19},
	doi = {10.1111/j.1469-8986.1982.tb02516.x},
	number = {6},
	journal = {Psychophysiology},
	author = {Dimberg, Ulf},
	year = {1982},
	pages = {643--647},
}

\end{document}